\documentclass[reprint,aps,prd,twocolumn,superscriptaddress,showpacs,preprintnumbers,amsmath,amssymb,floatfix,merge]{revtex4-1}
\usepackage[]{hyperref} 	
\usepackage[]{epsfig}
\usepackage{graphicx}
\usepackage{dcolumn}
\usepackage{epstopdf}
\usepackage{lineno}
\usepackage{multirow}

\newcommand{\irfu}{CEA, Centre d'Etudes Saclay, IRFU, 
91191 Gif-Sur-Yvette Cedex, France}
\newcommand{\ipnl}{IPNL, Universit\'{e} de Lyon, Universit\'{e} Lyon 1, 
CNRS/IN2P3, 4 rue E. Fermi 69622 Villeurbanne cedex, France}
\newcommand{\neel}{CNRS-N\'{e}el, 25 Avenue des Martyrs, 
38042 Grenoble cedex 9, France}
\newcommand{\csnsm}{CSNSM, Universit\'e Paris-Sud, IN2P3-CNRS, bat 108, 91405 Orsay,  France}
\newcommand{\iek}{Karlsruhe Institute of Technology,
Institut f\"{u}r Experimentelle Kernphysik, Gaedestr. 1, 76128 Karlsruhe, Germany}
\newcommand{\fzk}{Karlsruhe Institute of Technology,
Institut f\"ur Kernphysik, Postfach 3640, 76021 Karlsruhe, Germany}
\newcommand{\iramis}{CEA, Centre d'Etudes Saclay, 
IRAMIS, 91191 Gif-Sur-Yvette Cedex, France}
\newcommand{\jinr}{Laboratory of Nuclear Problems, JINR, Joliot-Curie 6, 
141980 Dubna, Moscow region, Russia}
\newcommand{\lsm}{Laboratoire Souterrain de Modane, CEA-CNRS, 
1125 route de Bardonn\`eche, 73500 Modane, France}
\newcommand{\oxford}{University of Oxford, Department of Physics, Keble Road, Oxford OX1 3RH, UK}
\newcommand{\sheffield}{Department of Physics and Astronomy, University of Sheffield, Hounsfield Road, Sheffield S3 7RH, UK}

\newcommand{\caltech}{Division of Physics, Mathematics \& Astronomy, California Institute of Technology, Pasadena, CA 91125, USA}
\newcommand{\cwr}{Department of Physics, Case Western Reserve University, Cleveland, OH  44106, USA}
\newcommand{\fnal}{Fermi National Accelerator Laboratory, Batavia, IL 60510, USA}
\newcommand{\lbnl}{Lawrence Berkeley National Laboratory, Berkeley, CA 94720, USA}
\newcommand{\massinttech}{Department of Physics, Massachusetts Institute of Technology, Cambridge, MA 02139, USA}
\newcommand{\queens}{Department of Physics, Queen's University, Kingston, ON, Canada, K7L 3N6}
\newcommand{\slac}{SLAC National Accelerator Laboratory/KIPAC, Menlo Park, CA 94025, USA}
\newcommand{\stolaf}{Department of Physics, St.\,Olaf College, Northfield, MN 55057 USA}
\newcommand{\santacla}{Department of Physics, Santa Clara University, Santa Clara, CA 95053, USA}
\newcommand{\smu}{Department of Physics, Southern Methodist University, Dallas, TX 75275, USA}
\newcommand{\stanford}{Department of Physics, Stanford University, Stanford, CA 94305, USA}
\newcommand{\syr}{Department of Physics, Syracuse University, Syracuse, NY 13244, USA}
\newcommand{\tam}{Department of Physics, Texas A \& M University, College Station, TX 77843, USA}
\newcommand{\ucb}{Department of Physics, University of California, Berkeley, CA 94720, USA}
\newcommand{\ucsb}{Department of Physics, University of California, Santa Barbara, CA 93106, USA}
\newcommand{\colo}{Departments of Phys. \& Elec. Engr., University of Colorado Denver, Denver, CO 80217, USA}
\newcommand{\ufl}{Department of Physics, University of Florida, Gainesville, FL 32611, USA}
\newcommand{\umn}{School of Physics \& Astronomy, University of Minnesota, Minneapolis, MN 55455, USA}
\newcommand{\zur}{Physics Institute, University of Z\"{u}rich, Winterthurerstr. 190, CH-8057, Switzerland}

\newcommand{\oi}{optimum interval}

\newcommand{\nicl}{$90\%~\mbox{CL}$}

\newcommand{\g}{\mbox{$\mbox{g}$}}
\newcommand{\kgd}{\mbox{$\mbox{kg} \! \cdot \! \mbox{d}$}}
\newcommand{\kgda}{\mbox{$\mbox{kg} \! \cdot \! \mbox{days}$}}
\newcommand{\keV}{\mbox{$\mbox{keV}$}}

\newcommand{\GeVcm}{\mbox{$\mbox{GeV/cm}^{3}$}}
\newcommand{\GeVc}{\mbox{$\mbox{GeV/c}^{2}$}}
\newcommand{\TeVc}{\mbox{$\mbox{TeV/c}^{2}$}}

\newcommand{\kms}{\mbox{$\mbox{km/s}$}}
\newcommand{\mwe}{\mbox{$\mbox{mwe}$}}
\newcommand{\cms}{\mbox{$\mbox{cm}^2$}}

\begin{document}

\title{Combined Limits on WIMPs from the CDMS and EDELWEISS Experiments
}

\affiliation{\caltech} 
\affiliation{\cwr}
\affiliation{\csnsm}
\affiliation{\fnal}
\affiliation{\ipnl}
\affiliation{\iramis}
\affiliation{\irfu}
\affiliation{\iek}
\affiliation{\fzk}
\affiliation{\jinr}
\affiliation{\lsm}
\affiliation{\lbnl}
\affiliation{\massinttech}
\affiliation{\neel}
\affiliation{\oxford}
\affiliation{\queens}
\affiliation{\sheffield}
\affiliation{\slac}
\affiliation{\stolaf}
\affiliation{\santacla}
\affiliation{\smu}
\affiliation{\stanford}
\affiliation{\syr}
\affiliation{\tam}
\affiliation{\ucb}
\affiliation{\ucsb}
\affiliation{\colo}
\affiliation{\ufl}
\affiliation{\umn}
\affiliation{\zur}

\author{Z.~Ahmed} \affiliation{\caltech} 
\author{D.~S.~Akerib} \affiliation{\cwr} 
\author{E.~Armengaud} \affiliation{\irfu}
\author{S.~Arrenberg} \affiliation{\zur}
\author{C.~Augier} \affiliation{\ipnl}
\author{C.~N.~Bailey} \affiliation{\cwr} 
\author{D.~Balakishiyeva} \affiliation{\ufl} 
\author{L.~Baudis} \affiliation{\zur}
\author{D.~A.~Bauer} \affiliation{\fnal} 
\author{A.~Beno\^{\i}t} \affiliation{\neel}
% ...

\collaboration{The CDMS and EDELWEISS Collaborations}
\noaffiliation

\date{\today}

\begin{abstract}
The CDMS and EDELWEISS collaborations have combined the results 
of their direct searches for dark matter using cryogenic germanium detectors.  
The total data set represents $614~\kgda$ equivalent exposure.
A straightforward method of combination was chosen for its simplicity before data were exchanged between experiments. 
The results are interpreted in terms of limits on spin-independent WIMP-nucleon  cross-section.
For a WIMP mass of $90~\GeVc$, where this analysis is most sensitive, a cross-section of $3.3 \times 10^{-44}~\cms$ is excluded at \nicl.
At higher WIMP masses, the combination improves the individual limits, by a factor  $1.6$ above $700~\GeVc$.
Alternative methods of combining the data  provide stronger constraints for some ranges of WIMP masses and weaker constraints for others.
\end{abstract}


\maketitle

The problem of dark matter has been an open question since 1933~\cite{Zwicky_1933}; 
most of the matter in the Universe appears only through its gravitational interactions. Evidence suggests that this dark matter may be made up of weakly interacting, massive particles (WIMPs)~\cite{bertone_particle_2010}.
Supersymmetric theories, and other extensions of the standard model of particle physics, predict plausible candidates for WIMPs~\cite{jungman_supersymmetric_1996,bergstroem_dark_2009}, and experimental efforts have been under way since the mid-1980s to detect them~\cite{spooner_direct_2007,bertone_particle_2010,schnee_introduction_2011}.   
The challenge is great because the average deposited energies involved are quite low, between a few $\keV$ and a few tens of $\keV$, 
and because the event rates are minute compared to normal levels of radioactive backgrounds.  
Therefore, experiments searching for WIMPs are typically located underground to reduce exposure to cosmic radiation, and they generally deploy detectors with some form of background identification.
For instance, cryogenic detectors operating at tens of millikelvins use a simultaneous measurement of phonons and charge to efficiently reject most of the dominant radioactive background (e.g. gamma particles) which interacts with electrons and,  for a given energy deposit, ionizes more than the nuclear recoils induced by WIMP scattering.
The experiments CDMS~\cite{CDMS_Science_2010,PhysRevLett.102.011301,PhysRevLett.96.011302,PhysRevD.72.052009} and EDELWEISS~\cite{EDW_2011,armengaud_first_2010,edelweiss_collaboration_final_2005}, 
have set strong constraints on WIMPs over the past decade using this technique with germanium targets.
Other experiments use other target nuclei and
techniques~\cite{xenon100_collaboration_dark_2011,angloher_commissioning_2009,zeplin2009,aubin_discrimination_2008,*ardm2010,*lux2010,*boulay_dark_2008,*cogent2011,*dama2010}.
Use of the same target nucleus and experimental technique by CDMS and EDELWEISS offers the possibility of combining their results to establish stronger constraints on WIMPs without introducing more model dependence 
than the individual results already had.

\begin{figure}[h]
	\centering
	\epsfig{file=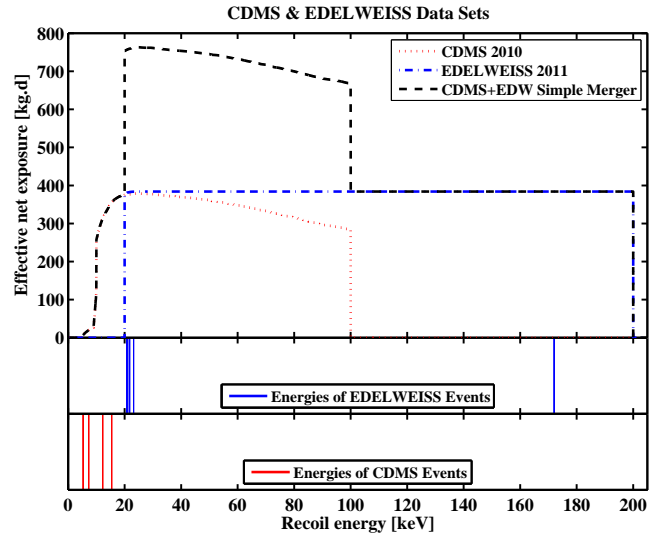,width=\linewidth}
	\caption[Exposure-weighted efficiencies]{Top: 
	effective net exposure
	curves for CDMS (dotted red curve), EDELWEISS (dash-dotted blue curve), and for the combined experiments (dashed black curve).  Middle: energies of the five observed candidate events in EDELWEISS (blue lines).  
	Bottom: energies of the four observed candidate events in CDMS (red lines).  All are below the EDELWEISS threshold.  In the simple method of merging the two sets of data, experimental provenance of the individual events is not kept.}
	\label{fig_expo}
\end{figure}

Published results~\cite{CDMS_Science_2010,PhysRevLett.102.011301,PhysRevLett.96.011302,PhysRevD.72.052009} from the CDMS experiment were obtained at the Soudan Underground Laboratory in Minnesota, a site of intermediate depth at $2100$~meters water equivalent (\mwe).  
CDMS has operated between six and thirty cryogenic semi-conductor detectors.  
In addition to the dual phonon-ionization measurement used to reject the majority of the highly-ionizing electron recoil background in the bulk of the detectors, CDMS devices measure non-thermal phonon information to identify background near the surface of the detector that could otherwise be confused with signal~\cite{akerib_surface_2007}.  
Moreover, all installed detectors have been used to identify backgrounds through multiple scattering.  
However, only the germanium ones (each of mass $\sim 230~\mbox{g}$) that were fully functioning (between four and fifteen at various times) were used to set  constraints on the WIMP cross section.
Thresholds as low as $5~\keV$ have been obtained in improved analyses~\cite{ogburn_PhD} of certain runs~\cite{PhysRevD.72.052009,PhysRevLett.96.011302}. 
The effective net exposure of the CDMS data set as a function of recoil energy, obtained by multiplying the energy-dependent efficiency by the exposure, is shown in Fig.~\ref{fig_expo}~\cite{supp_info}.
It reaches a maximum of $379~\kgd$ at  $25~\keV$.  
It then drops by roughly a quarter 
to the value obtained at $100~\keV$, the highest energy considered in these CDMS searches.
To avoid bias, these cuts were set in a blind manner, without knowledge of events in, or near, the signal region.
The four remaining candidate events observed in the CDMS data are all below an energy of $20~\keV$ 
(see Table~\ref{tab_energies} for energies of individual events).  
The expected number of background events misidentified as signal candidates in the full CDMS data set is $\sim 2$~events~\cite{CDMS_Science_2010,PhysRevLett.102.011301,ogburn_PhD}, with significant systematic uncertainty on the background at the lowest energies~\cite{ogburn_PhD}.
\begin{table}[ht]
\caption{Energies of the candidate events observed in each experiment~\cite{supp_info}.}
\centering
\begin{ruledtabular}
\begin{tabular}{cccccccccc}
\multicolumn{10}{c}{\textrm{Event energy $\left[ \keV \right]$}} \\
\multicolumn{4}{c}{\textrm{CDMS}} &\hspace{10 mm} & \multicolumn{5}{c}{\textrm{EDELWEISS}}\\
\hline
 5.3 & 7.3 & 12.3 & 15.5 & \hspace{10 mm} & 20.8 & 21.1 & 21.8 & 23.2 & 172\\
\end{tabular}
\end{ruledtabular}
\label{tab_energies}
\end{table}
\begin{figure}[h]
	\centering
	\epsfig{file=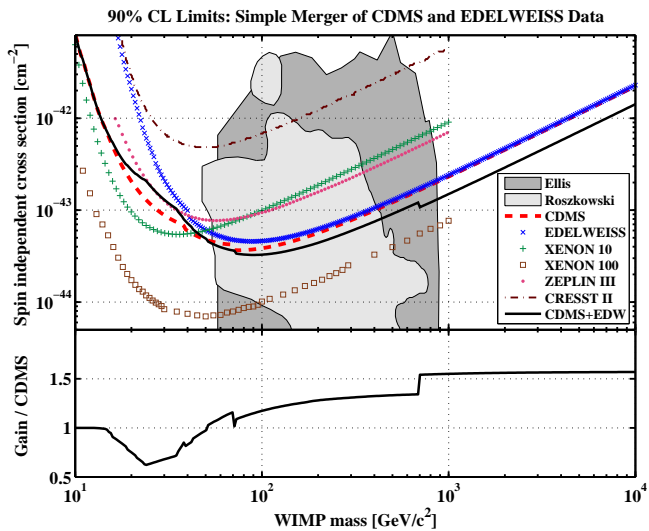,width=\linewidth}
	\caption[Main limits]{Top:  \nicl\  \oi\ upper limits on spin-independent WIMP couplings to nucleons as a function of WIMP mass, from the individual CDMS~\cite{CDMS_Science_2010} (red dashes)  and EDELWEISS~\cite{EDW_2011} (EDW, blue crosses) experiments, and from their simple merger (continuous black line).  
	Also represented are limits from the XENON~100~\cite{xenon100_collaboration_dark_2011} (brown boxes),
	XENON~10~\cite{xenon10_update} (green crosses),
	CRESST~II~\cite{angloher_commissioning_2009}(brown dot-dashed line) and ZEPLIN~III~\cite{zeplin2009} (pink dots)  experiments\footnote{The limits for XENON~10, CRESST~II and ZEPLIN~III use slightly larger halo escape velocities than this work ($600$--$650~\kms$).}, and supersymmetric parameter-space
	predictions~\cite{PhysRevD.71.095007,1126-6708-2007-07-075} (filled gray regions).  Bottom: gain obtained from the straightforward combination with respect to the strongest individual limit of CDMS and EDELWEISS (in effect that of CDMS).  Below masses of $50~\GeVc$, the combined limit is weaker than the best individual one; at higher masses, the gain is up to a factor $1.57$.}
	\label{fig_main_lim}
\end{figure}

The EDELWEISS experiment~\cite{EDW_2011} has deployed ten germanium detectors of 400~$\g$ at the Modane Underground Laboratory, a deeper site with a rock overburden of $4800~\mwe$. 
These detectors also use phonons and ionization to reject bulk background.  
In contrast to CDMS, the EDELWEISS detectors use patterned electrodes creating an inhomogeneous electrical field to identify surface background~\cite{EDW_ID}.  
The corresponding effective net exposure as a function of recoil energy is shown in Fig.~\ref{fig_expo}~\cite{supp_info}.
The distribution is very close to a step function with a sharp threshold of $20~\keV$ 
and a flat plateau at $384~\kgd$ up to $200~\keV$. 
Though not blind, cuts for the EDELWEISS data were set without considering events in the signal region.
Four of the candidate events observed in the EDELWEISS data set have energies in the $20$--$25~\keV$ range; the fifth event lies at $172~\keV$.  Precise values of energies are listed in Table~\ref{tab_energies}. 
The expected contribution of known backgrounds to these events is at most 3 events~\cite{EDW_2011}.  
Since the statistical sample is small and uncertainties in the backgrounds are still under investigation, all events are considered WIMP candidates for the purpose of deriving an upper limit on WIMPs.
Limits obtained by EDELWEISS are very similar to those of CDMS at high WIMP masses, though 
weaker at low WIMP masses.

Despite similar maximal exposures, the zero-background sensitivities of CDMS and EDELWEISS, estimated from the exposure-weighted efficiencies alone, differ significantly due to the higher threshold of the latter.
For example, the $10$--$100~\keV$ spectrum-averaged equivalent exposure for a WIMP of mass $90~\GeVc$ is $361~\kgd$ for CDMS
and $253~\kgd$ for EDELWEISS.
The corresponding values calculated for a mass of $1~\TeVc$ are $357~\kgd$ and $309~\kgd$, respectively.
They converge as the recoil energy spectrum hardens, explaining in large part the similar limits obtained at high mass.
The rest of the difference between the limits obtained by the two experiments 
is due to the number and energies of the nuclear-recoil candidates observed in each.  
Previously published limits 
from each experiment were calculated individually using Yellin's optimum interval 
method~\cite{yellin_2002}, which effectively sets a limit using the region of the  observed spectrum that is most constraining for the given model, and applies the appropriate statistical penalty for choosing this region.
All events are considered as potentially valid WIMP candidates; no expected background is subtracted.
This technique has the advantage that it does not require any a priori information on the backgrounds, and is well suited to cases with significant systematic uncertainties.

There are many possible procedures for combining results from experiments based on the \oi\ method~\cite{yellin_2011}.  Without some assumption about the backgrounds in the two experiments, it is impossible to determine which method has the best expected sensitivity.
Because not all background estimates for the full datasets considered had been made ``blindly'' before the signal regions were examined, and because of the uncertainties associated with these estimates, the collaborations decided not to try determining which method has the best expected sensitivity, but instead to use the same method each individual experiment had previously used~\cite{CDMS_Science_2010,EDW_2011} for combining data from different detectors and different data sets.  
In this simple merging method~\cite{yellin_2011}, the exposure-weighted efficiencies of both experiments are simply summed, and the events all treated on equal footing without consideration of their experiment of origin.  The experimental limits are then derived from the combined data set using the \oi\ method.
The decision  to use this method was made before data were exchanged, and indeed before the final analysis of the EDELWEISS data was available.

The combined effective net exposure curve is shown in Fig.~\ref{fig_expo}. 
The derived \nicl\ limit on the spin-independent cross-section for WIMP-nucleon scattering,  as a function of mass,  is shown in Fig.~\ref{fig_main_lim}.  We have assumed a standard astrophysical halo model (WIMP mass density $0.3~\GeVcm$, 
most probable WIMP velocity with respect to the galaxy $220~\kms$, 
mean velocity of  Earth with respect to the galaxy $232~\kms$, 
galactic escape velocity $544~\kms$~\cite{smith_rave_2007}), 
and the Helm form factor~\cite{lewin_review_1996}.
The limit is strongest at a mass of $90~\GeVc$ with a value of $3.3\times 10^{-44}~\cms$; it increases by less than $4\%$ over the $75$ to $115~\GeVc$ range.
Figure~\ref{fig_main_lim} also compares this limit with the results from the individual experiments.
The combined limits are stronger than those from the more sensitive of the two individual experiments (CDMS) for WIMP masses greater than $50~\GeVc$. 
The gain is as much as a factor $1.57$ at the highest masses, compared to a maximum allowed gain of $1.9$ from the relative experimental exposures alone. 
In the \oi\ technique, the gain results from the large eventless interval between $23.2$ and $172~\keV$.
Below a mass of $35~\GeVc$, the average recoil energy expected for a WIMP signal falls below $12~\keV$, and events start to appear in the optimum interval, further degrading the limit.
At high masses, the optimum interval contains no events, and the abrupt change in the limit is due to subtleties of the optimum 
interval method. 
Overall, the upper limits correspond to an expected number of detected WIMPs ranging from roughly 12 at low masses to about 4 at high masses.

\begin{figure}[h]
	\centering
	\epsfig{file=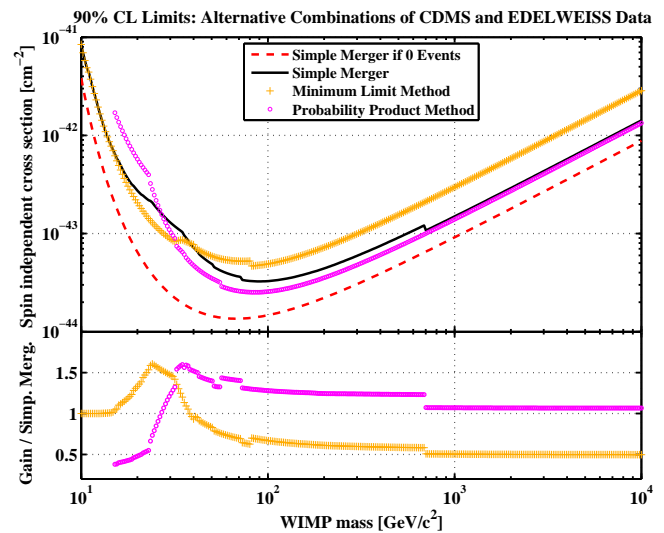,width=\linewidth}
	\caption[Other limits]{Comparison of combined CDMS-EDELWEISS limits obtained by different statistical methods.  The ``minimum limit" method (orange crosses) is more constraining than the adopted, ``simple merging" method (black curve) for low WIMP masses, but less constraining at high masses.  The ``probability product" method (purple circles) yields the strongest limits for masses above $30~\GeVc$, but has the weakest limits for low masses, 
	and is not defined for WIMPs with masses below $\sim 12~\GeVc$ since such WIMPs cannot be observed in EDELWEISS due to its the higher threshold. 
	The limit of the combined experiments under the simple method had no events been detected (red dashes) lies below these limits by a factor from $\sim 1.5$ to $\sim 9$. Bottom: gains of the alternative methods relative to the simple merger. 
	}
	\label{fig_other_meth}
\end{figure}

While this combination procedure does not preserve the identification of the candidate by its experiment of origin, and thus cannot take account of differing backgrounds, other methods can. Such methods could be of interest in future combinations of data sets with significantly different expected backgrounds, if those are known a priori. For comparison, Fig.~\ref{fig_other_meth} shows the results of two alternate procedures~\cite{yellin_2011} that may be expected to yield more constraining combined limits than the simple method used here.

The first alternate procedure (the ``minimum limit'' method~\cite{yellin_2011}) in essence computes the upper limit from each experiment separately, takes the lower of the two limits, and applies the appropriate statistical penalty for the choice.  
This method is suitable if both experiments are background-limited.
Figure~\ref{fig_other_meth} illustrates that this method results in less sensitive limits for the case at hand for high WIMP masses, since it effectively uses the results of only one experiment even for masses for which both effectively have negligible candidate events.  For low WIMP masses, however, it provides more stringent limits than the simple method.  This result also is to be expected: for masses for which the number of candidate events is large for both experiments, this method effectively determines which experiment provides the better limits and in practice ignores the other.

The second alternate procedure (the ``probability product" method~\cite{yellin_2011}) 
considers, at each WIMP mass, the product of the probabilities that each experiment excludes a given cross section as too high.
This method has poor sensitivity if experiments are background-limited, since an experiment with higher background will be weighted equally, worsening the limit of the more sensitive experiment.
It may have improved sensitivity compared to the simple method if both experiments are exposure-limited, yet have differing backgrounds.
As illustrated in Fig.~\ref{fig_other_meth}, this method gives a weaker limit than simple merging at low masses.
However, as  Fig.~\ref{fig_other_meth} also shows, for masses above $25~\GeVc$,
this method provides more stringent limits than the simple merger.  This result occurs because the optimum intervals for CDMS and EDELWEISS are different at these masses, so the extra freedom (compared to the simple method) to choose different intervals for the two experiments results in a greater sensitivity.
We note lastly that this procedure is not defined for WIMPs too light for at least one of the experiments to observe, as is the case for EDELWEISS and WIMPs below $\sim 12~\GeVc$.

The difference between the result of the minimum limit procedure and that of the probability product calculation suggests that the comparison of the two data sets could yield information on the backgrounds of each experiment, 
independent of that determined earlier~\cite{CDMS_Science_2010,EDW_2011}.
For each experiment, a likelihood was calculated as a function of WIMP mass using the experimental energy distribution of events and the expected distribution of WIMP events, and assuming no background. The EDELWEISS event at $172~\keV$ is ignored since, even for a heavy $10~\TeVc$ WIMP, the probability that at least one of the five EDELWEISS events occurs at $172~\keV$ or above is less than $2\%$.  
For CDMS and EDELWEISS individually, the maximum likelihood is obtained for WIMP masses below $17~\GeVc$; however, for such masses, the expected rate in CDMS is much higher than that in EDELWEISS, in contradiction with the measurement.
To further investigate compatibility, we performed a likelihood ratio test.
It rejects the hypothesis of no background at a confidence level greater than $99.8\%$.  
This result is fairly insensitive to changes in the WIMP astrophysical distribution, for instance increasing the escape velocity up to $650~\kms$, or varying the average WIMP velocity between $150$ and $350~\kms$.
Note however that no background subtraction has been performed to establish the limits in this work.

In conclusion, we have improved constraints from 
subkelvin germanium
detectors on WIMPs of mass greater than $\sim 50~\GeVc$ by a factor of up to $\sim 1.6$ thanks to a simple merger of the data from the CDMS and EDELWEISS experiments.  
Alternative methods that exploit the provenance of events could provide even stronger constraints at certain masses.  
Except for simple merging, these methods apply without change to combining results from experiments with different target nuclei for a given set of astrophysical and physical assumptions.
For simple merging of experiments with different recoil energy scales due to different target nuclei, it is recommended~\cite{yellin_2011} that before merging, the recoil energy of each event be transformed into its experiment's cumulative detection probability function at that event's energy.

This work is supported in part by 
the National Science Foundation (Grant Nos. AST-9978911, PHY-0542066, PHY-0503729, PHY-0503629, PHY-0503641, PHY-0504224, PHY-0705052, PHY-0801708, PHY-0801712, PHY-0802575 and PHY-0855525), 
by NSERC Canada (Grant SAPPJ~386399),
by the Department of Energy (Contracts DE-AC03-76SF00098, DE-FG02-91ER40688, DE-
FG02-92ER40701, DE-FG03-90ER40569, and DE-FG03-91ER40618), 
by the Swiss National Foundation (SNF Grant No. 20-118119)
by the Agence Nationale pour la Recherche under contract ANR-06-BLAN-0376-01, 
by the Russian Foundation for Basic Research,
and by the Science and Technology Facilities Council, UK.

\end{document}